\documentclass[aps, showpacs]{revtex4}
\usepackage{graphics}
\usepackage{epsfig}

\def\vb#1{\mbox{\boldmath$#1$}}
\def\pd#1#2{\frac{\partial #1}{\partial #2}}

\def\bdot{\,\vb{\cdot}\,}

\newcommand{\bc}{\begin{center}}
\newcommand{\ec}{\end{center}}
\newcommand{\bt}{\begin{tabbing}}
\newcommand{\et}{\end{tabbing}} 
\newcommand{\be}{\begin{eqnarray*}}
\newcommand{\ee}{\end{eqnarray*}}
\newcommand{\bs}{\begin{slide}}
\newcommand{\es}{\end{slide}}

\newcommand{\p}{\ensuremath{\mathbf{p}}}
\newcommand{\x}{\ensuremath{\mathbf{x}}}
\newcommand{\y}{\ensuremath{\mathbf{y}}}
\newcommand{\z}{\ensuremath{\mathbf{z}}}

\newcommand{\A}{\ensuremath{\mathbf{A}}}
\newcommand{\B}{\ensuremath{\mathbf{B}}}
\newcommand{\E}{\ensuremath{\mathbf{E}}}

\newcommand{\X}{\ensuremath{\mathbf{X}}}
\newcommand{\Y}{\ensuremath{\mathbf{Y}}}
\newcommand{\Z}{\ensuremath{\mathbf{Z}}}

\newcommand{\ahat}{\ensuremath{{\widehat{\mathbf{a}}}}}
\newcommand{\bhat}{\ensuremath{{\widehat{\mathbf{b}}}}}
\newcommand{\chat}{\ensuremath{{\widehat{\mathbf{c}}}}}

\newcommand{\ppar}{\ensuremath{{p_\|}}}
\newcommand{\pper}{\ensuremath{{p_\perp}}}

\newcommand{\ozero}{\ensuremath{{\mathcal{O}(1)}}}
\newcommand{\oone}{\ensuremath{{\mathcal{O}(\epsilon)}}}
\newcommand{\otwo}{\ensuremath{{\mathcal{O}(\epsilon^2)}}}
\newcommand{\As}{\ensuremath{{\A^*}}}
\newcommand{\Bs}{\ensuremath{{\B^*}}}
\newcommand{\Es}{\ensuremath{{\E^*}}}
\newcommand{\Phis}{\ensuremath{{\Phi^*}}}
\newcommand{\Bsp}{\ensuremath{{B^*_\|}}}

\newcommand{\ksf}{\ensuremath{\mathsf{k}}}
\newcommand{\psf}{\ensuremath{\mathsf{p}}}
\newcommand{\bsf}{\ensuremath{\mathsf{b}}}
\newcommand{\gsf}{\ensuremath{\mathsf{g}}}
\newcommand{\dsf}{\ensuremath{\mathsf{d}}}

\newcommand{\grad}{\ensuremath{\nabla}}

\newcommand{\curl}{\ensuremath{\nabla\!\times}}
\newcommand{\abs}[1]{\ensuremath{|#1|}}
\newcommand{\pp}[1]{\ensuremath{\frac{\partial}{\partial #1}}} 
\newcommand{\td}[2]{\ensuremath{\frac{d#1}{d#2}}}     
\newcommand{\tdp}[1]{\ensuremath{\frac{d}{d#1}}}  

\usepackage{amsmath}

\begin{document}            

\title{Hamiltonian theory of adiabatic motion of relativistic charged particles}
\author{Xin Tao} \author{Anthony A.~Chan}
\affiliation{Department of Physics and Astronomy, Rice University, Houston, Texas
77005}

\author{Alain J.~Brizard}
\affiliation{Department of Chemistry and Physics, Saint Michael's College, Colchester, Vermont 05439}

\begin{abstract}
A general Hamiltonian theory for the adiabatic motion of relativistic charged particles confined by slowly-varying background electromagnetic fields is presented based on a unified Lie-transform perturbation analysis in extended phase space (which includes energy and time as independent coordinates) for all three adiabatic invariants. First, the guiding-center equations of motion for a relativistic particle are derived from the particle Lagrangian. Covariant aspects of the resulting relativistic guiding-center equations of motion are discussed and contrasted with previous works. Next, the second and third invariants for the bounce motion and drift motion, respectively, are obtained by successively removing the bounce phase and the drift phase from the guiding-center Lagrangian. First-order corrections to the second and third adiabatic invariants for a relativistic particle are derived. These results simplify and generalize previous works to all three adiabatic motions of relativistic magnetically-trapped particles.
\end{abstract}


\maketitle

\section{Introduction}
The concept of the adiabatic motion of a charged particle in magnetic fields is important to research in space plasma physics and fusion physics \cite{Northrop1963, Brizard2007}. Depending on the confining magnetic geometry, a particle may display three quasi-periodic or periodic motions. The fastest of these three motions is the gyromotion about a magnetic field line (with frequency $\omega_\mathsf{g}$).  The second motion exists when a particle bounces along a magnetic field line between two mirror points (with frequency $\omega_\mathsf{b}$), because of nonuniformity along magnetic field lines. The slowest motion is the drift motion across magnetic field lines (with frequency $\omega_\dsf$) caused by perpendicular magnetic gradient-curvature drifts. In space physics  (and especially in radiation-belt physics \cite{Schulz1974}), these frequencies are widely separated such that $\omega_\mathsf{g} : \omega_\mathsf{b} : \omega_\mathsf{d} \sim \epsilon^{-1}: 1 : \epsilon$, where $\epsilon \ll 1$ is a small dimensionless ordering parameter to be defined below. Associated with each periodic orbital motion, there exists a corresponding adiabatic invariant.  We use $\mu$, $J_{\mathsf{b}}$ and $J_{\mathsf{d}}$ for the three invariants constructed in this work; to be consistent, we may also use $J_\mathsf{g} = (mc/q)\mu$, where $m$ is the particle's rest mass and $q$ its charge. 

The theory of the adiabatic motion of charged particles in electromagnetic fields has been well developed by Northrop \cite{Northrop1963}. However, the non-Hamiltonian method used by Northrop resulted in dynamical equations that do not possess important conservation properties like energy conservation in static fields, because of the absence of higher-order terms from Northrop's equations. In later work, Littlejohn \cite{Littlejohn1983} used a noncanonical phase-space transformation method, based on Lie-transform perturbation analysis, to obtain the Hamiltonian formulation of guiding-center dynamics for nonrelativistic particles.  By asymptotically removing the dependence on the gyrophase, the first invariant $J_{{\rm g}} = (mc/q)\,\mu$ is obtained from the guiding-center Lagrangian by Noether's theorem.  The resulting Hamiltonian guiding-center equations of motion conserve total energy for motion in static fields. In the present work, we use the Lie-transform perturbation analysis to develop a systematic Hamiltonian theory for relativistic guiding-center motion in weakly time-dependent electromagnetic fields. Our relativistic guiding-center equations of motion are expressed in semi-covariant form \cite{Boghosian1987}, which simplifies the previous work by Grebogi and Littlejohn \cite{Grebogi1984} (who extended their relativistic guiding-center equations to include ponderomotive effects associated with the presence of high-frequency electromagnetic waves) and generalizes earlier work by Brizard and Chan \cite{Brizard1999} (who considered guiding-center motion of a relativistic particle in static magnetic fields). 

The derivation of relativistic guiding-center dynamics begins with the removal of the gyrophase dependence from the particle phase-space Lagrangian. Since the condition for these periodic motions to exist is that the time variations of the forces a particle experiences should be slow compared to the particle's motion, we assume first that the electromagnetic fields vary on the drift timescale. Thus we shall construct the first and second adiabatic invariants from the particle's motion. While this ordering is not the most general case, it is the most common one in practice \cite{Littlejohn1983}. This procedure gives us the reduced six-dimensional guiding-center Lagrangian and the first invariant $J_\mathsf{g}$. Based on the guiding-center Lagrangian, we further remove the bounce phase and obtain the bounce-averaged guiding-center (or bounce-center) motion. The bounce-center Lagrangian for nonrelativistic particles has been derived by Littlejohn \cite{Littlejohn1982}, who at the same time constructs the second invariant $J_\mathsf{b}$ and the first-order correction to the second adiabatic invariant. The present work generalizes results of Littlejohn \cite{Littlejohn1982} and Brizard \cite{Brizard1990, Brizard2000} for relativistic particles. After we obtain the bounce-center Lagrangian, we change the time-scale ordering of the background fields so that the perturbation analysis can be applied to the drift motion.  We assume that the background fields vary on a time scale much slower than the drift time period when we derive the drift-center motion.  By drift averaging the bounce-center Lagrangian and removing the drift phase, we obtain the drift invariant $J_\mathsf{d}$ and the first-order correction to the third adiabatic invariant.

The remainder of the paper is organized as follows. In Sec.~\ref{sec:gc}, we derive the guiding-center equations of motion and the guiding-center Lagrangian for relativistic particles moving in slowly-varying electromagnetic fields. This Section generalizes previous work \cite{Littlejohn1983} for nonrelativistic particles and earlier work by Brizard and Chan \cite{Brizard1999} for relativistic particles moving in static magnetic fields only. In addition, by introducing effective covariant potentials, we also simplify the relativistic guiding-center equations of motion of Grebogi and Littlejohn \cite{Grebogi1984}. In Sec.~\ref{sec:bc}, we extend the work in Sec.~\ref{sec:gc} and use the Lie-transform method to obtain the bounce-center Lagrangian. The first-order correction to the second adiabatic invariant is automatically obtained in this process. In Sec.~\ref{sec:dc}, we assume that the electromagnetic fields vary on a time scale much slower than drift period and use a third Lie transform to remove the drift-phase dependence of the system and obtain the first-order correction to the third adiabatic invariant. A summary and comments on further work are given in Sec.~\ref{sec:summary}. 

\section{Relativistic guiding-center dynamics}
\label{sec:gc}

This Section presents the guiding-center equations of motion for a relativistic particle moving in slowly-varying background electromagnetic fields derived by the Lie-transform method. As the first step of the Lie transform, we show the ordering of the background fields, then we obtain the guiding-center Lagrangian which is later used to derive the guiding-center equations of motion and also to obtain the bounce-center Lagrangian in Sec.~\ref{sec:bc}.

\subsection{Background-field orderings}
Following the work of Littlejohn \cite{Littlejohn1983}, we use the small parameter $\epsilon\equiv\rho_0/L_0 \ll 1$ to order the background fields, where $\rho_0$ is the typical gyroradius and $L_0$ is the scale length of background fields. In dimensional units,  $\epsilon$ scales as $m /q$. We introduce the small parameter $\epsilon$ by denoting the physical electric field by $\E_{ph}$, and we assume that the $\E_{ph}\times\B$ drift is of $\oone$ compared to the particle's thermal speed \cite{Littlejohn1983,Grebogi1984}, and that the background fields $\E_{ph}$ and $\B$ vary on a time scale comparable to the drift period; i.e., $\partial/\partial t \sim \oone$. To indicate the order of a term explicitly by its $\epsilon$ factor, we set $\E_{ph}=\epsilon\E$, $\Phi_{ph} = \epsilon \Phi$ and $t_1 = \epsilon t$, where $\Phi_{ph}$ is the physical electric potential. Thus $\E\times\B\sim\ozero$, $\partial/\partial t_1 \sim \ozero$, and physical results are obtained by setting $\epsilon = 1$.

\subsection{Preliminary coordinate transformation}
With the ordering of background fields given above, the particle phase-space Lagrangian one-form \cite{Cary1983} in slowly-varying background electromagnetic fields is written in terms of extended (position, momentum; time, energy) phase-space coordinates $\z\,\equiv\,(\x, \p; t,  
W_\mathsf{p})$ as 
\begin{equation}
\label{pL1}  
\Gamma_\mathsf{p} \;=\; \left[\frac{1}{\epsilon}\frac{q}{c}\A(\x,t_1)+\p\right]\bdot d\x \;-\; W_\mathsf{p}\;dt \;-\; \mathcal{H}_\mathsf{p}\;d\sigma,
\end{equation}
where subscript `$\mathsf{p}$' denotes particle variables and $\mathcal{H}_\mathsf{p} = H_\mathsf{p} - W_\mathsf{p}$ is the extended particle Hamiltonian, with $H_\mathsf{p} = \gamma mc^2+q\Phi(\x,t_1)$ the Hamiltonian in regular phase space. Here, the physical dynamics takes place on the
surface $\mathcal{H}_\psf = 0$, the guiding-center relativistic factor is $\gamma\,\equiv\,\sqrt{1+\abs{\p/mc}^2}$, and $\sigma$ is an orbit  parameter. 

To show the dependence of $\Gamma_\mathsf{p}$ on the gyrophase explicitly, we decompose the relativistic momentum $\p$ according to
\begin{equation}
\label{lmc:def}
\p=\ppar_0\bhat+\sqrt{2m\mu_0 B}\,\chat,
\end{equation}
where $\ppar_0\equiv\p\bdot\bhat$ is the component of the relativistic momentum parallel to $\B$, $\mu_0\equiv\abs{\pper}^2/2mB$ will be shown to be the lowest-order term in the asymptotic expansion of an adiabatic invariant and $\chat$ is the perpendicular unit vector. The local momentum coordinates $(\ppar_0,\mu_0,\theta_0)$ \cite{Brizard1999} are then defined, where $\theta_0$ is the instantaneous gyrophase implicitly defined by the following relations \cite[]{Littlejohn1983}
\begin{equation}
\left. 
\begin{array}{rcl}
\ahat&=&\cos\theta_0\,\widehat{e}_1-\sin\theta_0\,\widehat{e}_2 \\
\chat&=&-\sin\theta_0\,\widehat{e}_1-\cos\theta_0\,\widehat{e}_2        
\end{array}
\right\},
\end{equation}
where $\chat$ is defined by Eq.~(\ref{lmc:def}), $\ahat=\bhat\times\chat$, and $(\widehat{e}_1,\widehat{e}_2,\widehat{e}_3)$ forms an arbitrary right-handed unit-vector set with $\widehat{e}_3\equiv\bhat$.

Substituting Eq.~(\ref{lmc:def}) into Eq.~(\ref{pL1}) yields the Lagrangian written in local momentum coordinates
\begin{eqnarray}
\label{pL2}
\Gamma_\mathsf{p} & = & \left[\frac{1}{\epsilon}\frac{q}{c}\A(\x,t_1)+\ppar_0\bhat+\sqrt{2m\mu_0 B}\,\chat\right]\bdot d\x \nonumber \\
 &  &-\; W_\mathsf{p}\,dt \;-\; \mathcal{H}_\mathsf{p}\,d\sigma.
\end{eqnarray}
Now $\gamma\equiv\sqrt{1+2\mu_0 B/mc^2+(\ppar_0/mc)^2}$ and $\Gamma_\mathsf{p}$ is a function of the preliminary phase-space coordinates $(\x,\ppar_0,\mu_0,\theta_0; t, W_\mathsf{p})$. Next, we will use a Lie transform to remove the gyrophase dependence from the particle Lagrangian (\ref{pL2}). 

\subsection{Guiding-center Lagrangian for a relativistic particle}
A Lie transform from the preliminary coordinates $\z=(\x, \ppar_0,\mu_0,\theta_0; t, W_\mathsf{p})$ to the guiding-center coordinates $\Z\equiv(\X, \ppar,\mu,\theta; t,  W_\mathsf{g})$ is used to remove the gyrophase dependence of $\Gamma_\mathsf{p}$. Here we use subscript '$\mathsf{g}$' to refer to guiding-center dynamics. For brevity, the steps of the guiding-center Lie transform are omitted here, but the interested reader may consult Ref.~\cite{Brizard1995} for details.  The resulting guiding-center Lagrangian in extended guiding-center phase-space coordinates $(\X, \ppar,\mu,\theta; t, W_\mathsf{g})$ is
\begin{eqnarray}
\label{gcL}
\Gamma_\mathsf{g} & = & \left[\frac{1}{\epsilon}\frac{q}{c}\A(\X,t_1)+\ppar\bhat(\X,t_1)+\oone\right]\bdot d\X \nonumber \\
 &  &+\; \epsilon\frac{mc}{q}\mu \,d\theta \;-\; W_\mathsf{g}\,dt \;-\; \mathcal{H}_\mathsf{g}\,d\sigma,
\end{eqnarray}
where the extended guiding-center Hamiltonian $\mathcal{H}_\mathsf{g}=H_\mathsf{g} - W_\mathsf{g}$, with the lowest-order regular Hamiltonian 
\begin{eqnarray}
H_\mathsf{g} & = & \gamma\,mc^2 \;+\; q\,\Phi(\X, t_1) \\
 & \equiv & mc^2\,\sqrt{1+2\mu B/mc^2 +(\ppar/mc)^2} \;+\; q\Phi(\X,t_1). \nonumber 
\end{eqnarray}
Here, the guiding-center coordinates are related to the preliminary coordinates and are given to lowest order in $\epsilon$ by $\ppar=\ppar_0$, $\mu =  \mu_0$, $\theta=\theta_0$, $W_\mathsf{g} = W_\mathsf{p}$, and
\begin{equation}
\label{GCCT} 
\X \;=\; \x-\epsilon\vb{\rho}, 
\end{equation}
where 
\begin{equation}
\vb{\rho}(\mu_0,\theta_0)\equiv\frac{c}{q}\sqrt{\frac{2m\mu_0}{B}}\,\ahat 
\end{equation}
is the gyroradius vector in guiding-center coordinates.  Note that because of the slow-time dependence, the differences between the guiding-center Lagrangian (\ref{gcL}) and that of Ref.~\cite{Brizard1999} are the electric potential and the time-changing variables, which give us extra second-order terms in the guiding-center equations of motion. 

\subsection{Guiding-center equations of motion}
Having found the relativistic guiding-center Lagrangian (\ref{gcL}), we now solve for guiding-center equations of motion using Euler-Lagrange equations \cite{Goldstein1980}. For a Lagrangian $\mathcal{L}_\mathsf{g}$, which is related to $\Gamma_\mathsf{g}$ in Eq.~(\ref{gcL}) by $\Gamma_\mathsf{g}\equiv \mathcal{L}_\mathsf{g}\,d\sigma$,  written in extended guiding-center phase-space coordinates $Z^\nu$, the Euler-Lagrange equation is 
\begin{equation}
\label{eq:6}
  \tdp{\sigma}\left(\pd{\mathcal{L_\mathsf{g}}}{\dot{Z}^{\nu}}\right) - \pd{\mathcal{L}_\mathsf{g}}{Z^{\nu}} = 0, 
\end{equation} 
where $\dot{Z}^\nu = dZ / d\sigma$. The equations of motion for $t$ and $W_\mathsf{g}$ are 
\begin{equation}
  \label{eq:5}
  \dot{t}=\td{t}{\sigma} = -\pd{\mathcal{H}_\gsf}{W_\mathsf{g}} = +1, 
\end{equation}
which indicates that $t$ and $\sigma$ can be identified, and the time rate change of energy
\begin{equation}
  \label{eq:7}
  \dot{W}_\mathsf{g} = q\pd{\Phis}{t_1} - \frac{q}{c}\dot{\X}\bdot\pd{\As}{t_1}, 
\end{equation}
where we replaced $\sigma$ by $t$ because of  Eq.~(\ref{eq:5}). Here, the effective potentials $\Phis$ and $\As$ are defined as 
\begin{equation}
  \label{eq:2}
  \begin{pmatrix}
    \Phis\\
    \epsilon^{-1}\As 
  \end{pmatrix}
  = 
  \begin{pmatrix}
    \Phi \\
    \epsilon^{-1}\A
  \end{pmatrix}
  \;+\; 
  \frac{mc}{q}
  \begin{pmatrix}
    \gamma c \\
    \gamma v_\|\bhat
  \end{pmatrix},
\end{equation}
where the second term on the right side represents the covariant {\it parallel two-flat} decomposition of the relativistic guiding-center four-velocity \cite{Boghosian1987}.

Applying the Euler-Lagrange equation (\ref{eq:6}) to other guiding-center phase-space coordinates $(\X, \ppar, \mu, \theta)$, we first have $\bhat\bdot\dot{\X}=\ppar/(\gamma\,m)$, showing the parallel motion of the guiding center; secondly, $\dot{\theta} = \epsilon^{-1}qB/(\gamma\,mc)$, showing the fast gyromotion, and $\dot{\mu} = 0$, which proves that $\mu$ is an invariant of the guiding-center motion (here, a dot means a total derivative with respect to $t$). Finally, the relativistic guiding-center equations for $\dot{\X}$ and $\dot{\ppar}$ are 
\begin{eqnarray}
\label{dX/dt}
\dot{\X}&=&\frac{\ppar}{\gamma m}\frac{\Bs}{\Bsp}+\epsilon\Es\times\frac{c\bhat}{\Bsp} \\
\label{dppar/dt}
\dot{\ppar}&=& q\Es\bdot\frac{\Bs}{\Bsp},
\end{eqnarray}
where the effective fields $(\Es, \Bs)$ are defined in terms of the potentials (\ref{eq:2}) as
\begin{equation}
  \label{def:Bs}
  \Bs\equiv\curl\As=\B+\epsilon\frac{c\ppar}{q}\curl\bhat,
\end{equation}
and
\begin{equation}
\label{def:Es}
\Es\equiv-\frac{1}{c}\pd{\As}{t_1}-\grad{\Phis}=\E -\frac{\epsilon}{q}\left(\ppar \pd{\bhat}{t_1}+mc^2\grad \gamma\right),
\end{equation}
where 
\begin{equation}
\label{def:Bsp}
  \Bsp\equiv\Bs\bdot\bhat=B+\epsilon(c\ppar/q)\bhat\bdot\curl\bhat,
\end{equation}
and $\nabla\gamma = (\mu/\gamma mc^{2})\,\nabla B$. Equation (\ref{dX/dt}) shows that the guiding-center velocity consists of the parallel motion along a field line, the $\E\times\B$, gradient-$B$ and curvature drifts. The curvature drift here is hidden in the first term on the right side of Eq.~(\ref{dX/dt}) and the gradient drift and the $\E\times\B$ drift are contained in the second term.  Equation (\ref{dppar/dt}) represents the parallel force along a field line, which according to Eq.~(\ref{def:Es}) consists of two parts: one from the parallel electric field and the other from the magnetic mirror force. Note that the first-order term in Eq.~(\ref{def:Es}) gives second-order terms in the guiding-center equation of motion, which are important to the conservation properties of the guiding-center motion. 

We immediately note the simplicity of the relativistic guiding-center equations of motion (\ref{dX/dt}) and (\ref{dppar/dt}), expressed in terms of the covariant effective potentials (\ref{eq:2}), compared to the relativistic guiding-center equations of motion of Grebogi and Littlejohn (GL) 
\cite{Grebogi1984}, who used the scalar potential $\Phi$ instead of the covariant potential $\Phi^{*}$. We recover the GL relativistic guiding-center equations of motion by substituting $q{\bf E}^{*} = q{\bf E}^{*}_{{\rm GL}} - \epsilon\,(\mu/\gamma)\,\nabla B$ in Eqs.~(\ref{dX/dt})-(\ref{dppar/dt}). We also point out that, in contrast to Boghosian's manifestly-covariant formulation for relativistic guiding-center motion \cite{Boghosian1987}, our ``$1 + 3$'' semi-covariant formulation treats time separately from the other phase-space coordinates and uses an energy-like Hamiltonian (instead of the Lorentz-invariant covariant Hamiltonian).

If the fields are static, then Eq.~(\ref{eq:7}) shows conservation of energy automatically. Also, the relativistic guiding-center Eqs.~(\ref{dX/dt})-(\ref{dppar/dt}) satisfy the Liouville theorem
\begin{equation}
  \label{eq:34}
  \pd{\Bsp}{t} + \grad\bdot\left(\Bsp\dot{\X}\right)+\pp{\ppar}{\left(\Bsp\dot{\ppar}\right)}=0,
\end{equation}
which ensures that guiding-center phase-space volume is conserved by the guiding-center dynamics. We prove Eq.~(\ref{eq:34}) explicitly as follows. First, we easily obtain from Eqs.~(\ref{dX/dt}) - (\ref{dppar/dt})
\begin{eqnarray}
  \label{eq:36}
  \pd{\Bsp}{t}&=&\bhat\bdot\pd{\Bs}{t} + \Bs\bdot\pd{\bhat}{t}, \\
  \label{eq:37}
  \grad\bdot\left(\Bsp\dot{\X}\right)&=& c\left(\bhat\bdot\curl\Es-\Es\bdot\curl\bhat\right) \nonumber \\
 &  &+\; \frac{\ppar}{m}\Bs\bdot\grad(\gamma^{-1}), \\
  \label{eq:38}
  \pp{\ppar}{\left(\Bsp\dot{\ppar}\right)}&=& q\left(\pd{\Es}{\ppar}\bdot\Bs+\Es\bdot\pd{\Bs}{\ppar}\right).
\end{eqnarray}
Next, we insert 
\begin{eqnarray}
  \pd{\Es}{\ppar}&=&-\frac{\ppar}{mq}\grad(\gamma^{-1})-\frac{1}{q}\pd{\bhat}{t}, \\
  \pd{\Bs}{\ppar}&=&\frac{c}{q}\curl\bhat, 
\end{eqnarray}
and 
\begin{equation}
  \label{eq:51}
  \pd{\Bs}{t} = -c\curl\Es
\end{equation}
in Eqs.~(\ref{eq:36})-(\ref{eq:38}) to find that Eq.~(\ref{eq:34}) is satisfied exactly, where we have set $\epsilon = 1$ in Eqs.~(\ref{eq:34})-(\ref{eq:51}). 

\section{Hamiltonian theory of bounce-center dynamics}
\label{sec:bc}
To obtain the bounce-center Lagrangian, we preform a Lie transform on the relativistic guiding-center Lagrangian (\ref{gcL}) to remove the bounce-phase dependence. This Lie transform leads to construction of the second adiabatic invariant and gives the first-order correction to the second adiabatic invariant directly. The nonrelativistic bounce-center Lagrangian has been derived by Littlejohn \cite{Littlejohn1982}, and the present work generalizes previous results to the relativistic case.   

\subsection{Preliminary coordinate transformation}
We first drop the term $\epsilon(mc/q)\mu d\theta$ in the extended guiding-center Lagrangian (\ref{gcL}), which means we are now considering a six-dimensional system parametrized by constant-$\mu$ surfaces. Also we separate the extended Hamiltonian $\mathcal{H}_\mathsf{g}$ from the symplectic part of the extended phase-space Lagrangian (\ref{gcL}) (as in Eq.~(\ref{eq:35}) below). We then perform a coordinate transformation from $\X$ to $(\alpha,\beta,s)$, where $(\alpha,\beta)$ are the usual Euler potentials such that $\B=\grad\alpha\times\grad\beta=B\bhat$,  and $s$ is the position along a field line labeled by $(\alpha,\beta)$, with $\bhat = \partial \X/\partial s$.  We choose the vector potential $\A = \alpha \grad \beta$, write 
\begin{equation}
  \label{eq:28}
  d\X\, = \, \pd{\X}{\alpha} d\alpha + \pd{\X}{\beta} d\beta + \bhat\, ds + \pd{\X}{t_1}dt_1, 
\end{equation}
and we write the (symplectic part of the) guiding-center Lagrangian (\ref{gcL}) order by order as
\begin{equation}
  \label{eq:35}
  \Gamma_\mathsf{g} = \frac{1}{\epsilon}\sum_{n=0}^{\infty}\epsilon^n\Gamma_{\mathsf{g}n},
\end{equation}
where
\begin{equation}
  \label{gcL0a}
  \Gamma_{\mathsf{g}0}\, =\, \frac{q}{c}\alpha d \beta - K_{\mathsf{g}}dt_1,
\end{equation}
and the modified guiding-center energy coordinate
\begin{equation}
\label{eq:3}
K_{\mathsf{g}} \equiv W_\mathsf{g} + \frac{q}{c}\alpha\pd{\beta}{t_1}.
\end{equation}
Eq.~(\ref{eq:3}) introduces a change to the extended Hamiltonian $\mathcal{H}_\mathsf{g} = H_\mathsf{g}-K_\mathsf{g}$, where the lowest-order ordinary Hamiltonian $H_\mathsf{g}$ is $H_{\mathsf{g}0} = q\Phis + (q/c)\alpha\,\partial\beta/ \partial t_1$. It is also useful to follow Littlejohn \cite{Littlejohn1982} by using a 2-vector $\y$ with $y_1\,=\,\alpha, y_2\,=\,\beta$, together with the two-dimensional Levi-Civita symbol $\eta_{ab}$, where $a, b$ runs overs 1 and 2. The components of $\eta_{ab}$ are given by $\eta_{11}\,=\,\eta_{22}=\,0$ and $\eta_{12}\,=\,-\eta_{21}\,=\,1$. 

Before considering the first-order term $\Gamma_{\mathsf{g}1}$ in Eq.~(\ref{eq:35}) written in coordinates $(\y,s)$, we make the usual assumption about the lowest-order motion that,  with coordinates $(\y, t_1)$ frozen, the bounce motion in $(s,\ppar)$ space is periodic \cite{Brizard2000}.  Thus, using the Hamilton-Jacobi theory \cite{Goldstein1980}, we construct the action-angle canonical variables $(J_{\mathsf{b}0}, \psi_{{\mathsf{b}0}})$ corresponding to the periodic bounce motion. Then  
\begin{equation}
  J_{\mathsf{b}0}(\alpha, \beta, \mu; t,  K_{\mathsf{g}})\,=\,\frac{1}{2\pi}\oint\ppar ds,
\end{equation}
and $\omega_{\mathsf{b}0}$ is the lowest-order angular bounce frequency, defined by $\omega_{\mathsf{b}0}^{-1}\,=\,\partial J_{\mathsf{b}0}/\partial {K_{\mathsf{g}}}$.  The bounce-phase angle $\psi_{\mathsf{b}0}$ is canonically conjugate to $J_{\mathsf{b}0}$. Also, the following relation holds for the true motion:
\begin{equation}
  \label{eq:24}
  \pd{s}{\psi_{\mathsf{b}0}}\pd{\ppar}{J_{\mathsf{b}0}}-\pd{s}{J_{\mathsf{b}0}}\pd{\ppar}{\psi_{\mathsf{b}0}}\,=\,1,
\end{equation}
since the transformation from $(\ppar, s)$ to $(J_{\mathsf{b}0}, \psi_{\mathsf{b}0})$ is canonical. At lowest order, $J_{\mathsf{b}0}$ is an invariant of motion. When higher-order terms are included and $(\y,t_1)$ are allowed to evolve, we will show that $\dot{J_{\mathsf{b}0}}\,=\,\oone$. The symmetry of the unperturbed motion has been pointed out by Littlejohn \cite{Littlejohn1983}, and we will directly use this result to simplify the expression of the first-order correction to the second adiabatic invariant. 

Using the coordinates $Z_{0}^{\mu} = (\y, J_{\bsf 0}, \psi_{\bsf 0}, t_1, W_{\bsf 0})$, the first order guiding-center Lagrangian in Eq.~(\ref{eq:35}) has the components 
\begin{equation}
  \label{eq:8}
   \Gamma_{\mathsf{g}1\mu}\,=\,\ppar\bhat\bdot\pd{\X}{Z_{0}^{\mu}} \;\equiv\; p_{\|}\,b_{\mu},
\end{equation}
and we will omit subscripts of $J_{\mathsf{b}0}$ and $\psi_{\mathsf{b}0}$ when they themselves are subscripts. Note that in covariant form, $\bhat = \grad s + b_a \grad y_a$. These expressions will be further simplified with the second coordinate transformation from $(\ppar,s)$ to $(J_{\mathsf{b}0},\psi_{\mathsf{b}0})$. 

\subsection{Coordinate Transformation from $(\ppar,s)$ to $(J_{\mathsf{b}0},\psi_{\mathsf{b}0})$}
To simplify components of $\Gamma_{\mathsf{g}1}$ defined in Eq.~(\ref{eq:8}), we perform a coordinate transformation from $(\ppar,s)$ to $(J_{\mathsf{b}0},\psi_{\mathsf{b}0})$ by adding a gauge term $dG_{\mathsf{b}}$ to $\Gamma_{\mathsf{g}1}$ (i.e., $\Gamma'_{\mathsf{g}1}\,=\,\Gamma_{\mathsf{g}1}+dG_\mathsf{b}$)  to eliminate the $J$-component ($\Gamma_{\mathsf{g}1J}' = 0$). Thus we choose $G_\bsf$ to be
\begin{equation}
  \label{eq:16}
  G_\mathsf{b}\,=\,-\int^{J_{\mathsf{b}0}}_0 dJ'\,\Gamma_{\mathsf{g}1J}(t_1,\psi_{\mathsf{b}0},J',y). 
\end{equation}
Also, the $\psi$-component $\Gamma'_{\mathsf{g}1\psi}=\Gamma_{\gsf1\psi}+\partial G_\bsf/\partial \psi_{\bsf 0}$ becomes 
\begin{equation}
  \label{eq:39}
  \Gamma'_{\mathsf{g}1\psi} \;=\; \int^{J_{\mathsf{b}0}}_0dJ' \left[ \pd{\left( \ppar\,b_{\psi} \right)}{J'} \;-\; \pd{\left( 
\ppar\,b_{J} \right)}{\psi_{\bsf 0}} \right] \;=\; J_{\mathsf{b}0},
\end{equation}
where we have used Eq.~(\ref{eq:24}) to get the second equality and used $\Gamma_{\mathsf{g}1J}\,=\,0$ at $J_{\mathsf{b}0}=0$ (since $\ppar = 0$). 

Similarly, the other components of $\Gamma'_{\mathsf{g}1}$ are
\begin{equation}
\label{eq:40}
\Gamma'_{\mathsf{g}1a}\,=\,\int^{J_{\mathsf{b}0}}_0 dJ'\;\left[ \pd{\left( \ppar\,b_{a} \right)}{J'} \;-\; \pd{\left( 
\ppar\,b_{J} \right)}{y_a} \right] \,\equiv\, F_a,
\end{equation}
and 
\begin{equation}
\label{eq:42}
\Gamma'_{\mathsf{g}1t_1}\,=\,\int^{J_{\mathsf{b}0}}_0 dJ'\;\left[ \pd{\left( \ppar\,b_{t} \right)}{J'} \;-\; \pd{\left( 
\ppar\,b_{J} \right)}{t_1} \right] \,\equiv\, F_{t_1}.
\end{equation}
Combining Eqs.~(\ref{eq:39}) - (\ref{eq:42}), the new first-order guiding-center Lagrangian is 
\begin{equation}
\Gamma_{\mathsf{g}1}'\,=\,F_a\,dy_a+J_{\mathsf{b}0}\,d\psi_{\mathsf{b}0}+F_{t_1} dt_1.
\end{equation}
The Euler-Lagrange equation [see Eq.~(\ref{eq:6})] for $J_{\bsf 0}$ obtained from $\Gamma_{\gsf}' = \Gamma_{\gsf 0} + \epsilon\Gamma_{\gsf 1}'\equiv \mathcal{L}_\gsf' d\sigma$ is 
\begin{equation}
  \td{J_{\mathsf{b}0}}{t}\,=\,\pd{F_a}{\psi_{\mathsf{b}0}}\dot{y}_a+\epsilon\,\pd{F_{t_1}}{\psi_{\mathsf{b}0}}+\otwo,
\end{equation}
where to lowest order
\begin{equation}
\dot{y}_a = \epsilon \frac{c}{q}\eta_{ab}\pd{H_{\mathsf{g}0}}{y_b}.
\end{equation}
 
For later use, we now write the first two terms of the guiding-center Lagrangian [Eq.~(\ref{eq:35})] in coordinates $(\alpha, \beta, J_{\mathsf{b}0}, \psi_{\mathsf{b}0}; t_1, K_\mathsf{g})$ with the prime dropped, 
\begin{eqnarray}
\label{eq:21}
    \Gamma_{\mathsf{g}0}& =& \frac{q}{c}\alpha d \beta - K_{\mathsf{g}}dt_1,  \\
\label{eq:22}
    \Gamma_{\mathsf{g}1}&=&F_ady_a+J_{\mathsf{b}0}d\psi_{\mathsf{b}0}+F_{t_1} dt_1,
\end{eqnarray}
and the lowest order Hamiltonian is 
\begin{equation}
\label{eq:23}
\mathcal{H}_\mathsf{g}=H_{\mathsf{g}0} - K_{\mathsf{g}}. 
\end{equation}

With these coordinate transformations and Lagrangian, we do a Lie transform to remove the bounce-phase dependence from $\Gamma_{\mathsf{g}}$ and obtain the bounce-center Lagrangian $\Gamma_\mathsf{b}$. 

\subsection{Lie transform in extended phase-space coordinates}
The bounce-center dynamics are obtained using the Lie transform in extended phase-space coordinates.  The bounce-center Lagrangian and Hamiltonian are constructed order by order
\begin{eqnarray}
  \label{eq:43}
  \Gamma_{\mathsf{b}}&=&\Gamma_{\mathsf{b}0}+\epsilon\Gamma_{\mathsf{b}1}+\epsilon^2\Gamma_{\mathsf{b}2}+\cdots, \\
  \label{eq:44}
  \mathcal{H}_\mathsf{b}&=&\mathcal{H}_{\mathsf{b}0}+\epsilon\mathcal{H}_{\mathsf{b}1}+\epsilon^2\mathcal{H}_{\mathsf{b}2}+\cdots,
\end{eqnarray} 
where the terms on the right-hand side of Eq.~(\ref{eq:43}) are
\begin{eqnarray}
  \label{eq:46}
  \Gamma_{\mathsf{b}0}&=&\Gamma_{\mathsf{g}0}, \\
  \label{eq:4}
  \Gamma_{\mathsf{b}1}&=&\Gamma_{\mathsf{g}1}-i_1\cdot\Omega_{\mathsf{g}0}+dS_1, \\
  \label{eq:45}
  \Gamma_{\mathsf{b}2}&=&-i_2\cdot\Omega_{\mathsf{g}0}-i_1\cdot\Omega_{\mathsf{g}1} \nonumber \\
 &  &\;+\frac{i_1}{2}\cdot d(i_1\cdot\Omega_{\mathsf{g}0}) + dS_2,
\end{eqnarray}
and the first two terms in Eq.~(\ref{eq:44}) are
\begin{eqnarray}
  \mathcal{H}_{\mathsf{b}0}&=&\mathcal{H}_{\mathsf{g}0}, \\ 
  \label{eq:25}
  \mathcal{H}_{\mathsf{b}1}&=&\mathcal{H}_{\mathsf{g}1} - g_1\cdot d\mathcal{H}_{\mathsf{g}0}. 
\end{eqnarray}
The term $i_n\cdot\Omega_{\mathsf{g}} = g_n^\mu (\Omega_{\mathsf{g}})_{\mu\nu} dZ^\nu$ in Eqs.~(\ref{eq:4}) - (\ref{eq:45}) and the term $g_n\cdot d\mathcal{H}_\mathsf{g} = g_n^\mu\cdot \partial \mathcal{H}_\mathsf{g} / \partial Z^\mu$ in Eq.~(\ref{eq:25}) are expressed in terms of the $n^{th}$-order Lie-transform generating vector $g_n$ and gauge function $S_n$, where 
\begin{equation}
  (\Omega_{\mathsf{g}})_{\mu\nu}\equiv[Z^\mu, Z^\nu]= \pd{\Gamma_{\mathsf{g}\nu}}{Z^\mu}-\pd{\Gamma_{{\mathsf{g}}\mu}}{Z^\nu}
\end{equation}
is the Lagrange bracket between $Z^\mu$ and $Z^\nu$. 

\subsection{Bounce-center motion in coordinates $(\Y, J_\mathsf{b},\psi_\mathsf{b}; t, K_\mathsf{b})$}
\label{sec:bounce-center-motion}
Following the Lie-transform procedure described in Eqs.~(\ref{eq:46})-(\ref{eq:25}), we first have the lowest-order Lagrangian and Hamiltonian 
\begin{eqnarray}
  \Gamma_{\mathsf{b}0}& =& \frac{q}{c}Y_\alpha d Y_\beta - K_{\mathsf{b}}dt_1, \\
  \mathcal{H}_{\mathsf{b}0} & = & H_{\mathsf{b}0} - K_{\mathsf{b}},
\end{eqnarray}
where $(Y_\alpha, Y_\beta)$ represent the bounce-center coordinates $\Y$, and $H_{\mathsf{b}0}$ has the same functional dependence on the bounce-center coordinates $(\Y, J_\mathsf{b}, t)$ as $H_{\mathsf{g}0}$ on the guiding-center coordinates $(\y, J_\mathsf{b0}, t)$. 

The first-order bounce-center Lagrangian (\ref{eq:4}) then becomes  
\begin{equation}
  \label{eq:31}
  \Gamma_{\mathsf{b}1} = (-g_1^b\Omega_{ab0} + F_a) dy_a + J_{\mathsf{b}0}\, d\psi_{\mathsf{b}0} + (F_{t_1} + g_1^K) dt_1,
\end{equation}
where $\Omega_{ab0} = -(q/c)\eta_{ab}$ and we choose $S_1 = 0$ in Eq.~(\ref{eq:4}). Requiring $\Gamma_{\mathsf{b}1a}=0$ and $\Gamma_{\mathsf{b}1t_1} = 0$ gives us  
\begin{eqnarray}
  \label{eq:10}
  g^a_1 &=& -\frac{c}{q}\eta_{ab}F_b, \\
  \label{eq:11}
  g^K_1 &=& -F_{t_1}.
\end{eqnarray}
The first-order Hamiltonian then is given by
\begin{equation}
  \label{eq:32}
  \mathcal{H}_{\mathsf{b}1}=-g_1^a\pd{H_{\mathsf{g}0}}{y_a}-g_1^J\omega_{\mathsf{b}0}+g_1^K,
\end{equation}
since $H_{\mathsf{g}1} = 0$ in Eq.~(\ref{eq:25}). With Eqs.~(\ref{eq:10}), (\ref{eq:11}), and the requirement $\widetilde{\mathcal{H}}_{\mathsf{b}}=0$, where a tilde in this section denotes the bounce-phase oscillatory part, we have
\begin{equation}
\label{eq:30}
  \widetilde{g_1^J} = \frac{1}{\omega_{\mathsf{b}0}}\left(-\widetilde{F}_{t_1} + \frac{c}{q}\eta_{ab}\widetilde{F}_b\pd{H_{\mathsf{g}0}}{y_a}\right),
\end{equation} 
and 
\begin{equation}
  \label{eq:29}
  \mathcal{H}_{\mathsf{b}1}=\frac{c}{q}\eta_{ab} \langle F_b \rangle\pd{H_{\mathsf{g}0}}{y_a}-\langle g_1^J\rangle\omega_{\mathsf{b}0}-\langle F_{t_1}\rangle,
\end{equation}
where $\langle\cdots\rangle$ denotes a bounce-phase average. It has been shown in Ref.~ \cite{Littlejohn1982} that $F_a$ and $F_{t_1}$ are odd in $\psi_{\bsf0}$, and thus we have $\langle F_a \rangle = 0$ and $\langle F_{t_1}\rangle = 0$.  Equation (\ref{eq:29}) then becomes 
\begin{equation}
  \label{eq:27}
  \mathcal{H}_{\mathsf{b}1}=-\langle g_1^J\rangle\omega_{\mathsf{b}0}.
\end{equation}

To obtain $g_1^\psi$ and the bounce-phase averaged part of $g_1^J$ needed in Eq.~(\ref{eq:27}), we need to go to the second-order Lie transform of the  Lagrangian.  The $\psi_{\mathsf{b}0}$ part and the $J_{\mathsf{b}0}$ part of the second order Lagrangian $\Gamma_{\mathsf{b}2}$ are
\begin{eqnarray}
  \label{eq:12}
  \Gamma_{\mathsf{b}2\psi} &=& \pd{S_2}{\psi_{\mathsf{b}0}}-g_1^J-\frac{1}{2}\frac{c}{q}\eta_{ab}F_b\pd{F_a}{\psi_{\mathsf{b}0}}, \\
  \label{eq:13}
  \Gamma_{\mathsf{b}2J} &=& \pd{S_2}{J_{\mathsf{b}0}}+g_1^\psi-\frac{1}{2}\frac{c}{q}\eta_{ab}F_b\pd{F_a}{J_{\mathsf{b}0}}.
\end{eqnarray}
 To make $J_{\mathsf{b}}$ the exact invariant, we require that $\Gamma_{\mathsf{b}2\psi} = 0$. Taking the bounce-averaged part of Eq.~(\ref{eq:12}) and using $\langle S_2\rangle= 0$, we have $\langle g_1^J\rangle = 0$, since $\partial F_a /\partial \psi_{\bsf 0}$ is even in $\psi_{b0}$, and thus $\langle F_b\,\partial F_a/\partial \psi_{\mathsf{b}0}\rangle = 0$. This result indicates that Eq.~(\ref{eq:27}) becomes $\mathcal{H}_{\mathsf{b}1} = 0$. 

The bounce-phase dependent part of Eq.~(\ref{eq:12}) is solved to give the gauge function
\begin{equation}
\label{eq:14}
S_2  = \int d\psi_{\mathsf{b}0}\left(g^J_1+\frac{1}{2}\frac{c}{q}\,\eta_{ab}F_b\pd{F_a}{\psi_{\mathsf{b}0}}\right),  
\end{equation}
where $g^J_1 = \widetilde{g^J_1}$ is given in Eq.~(\ref{eq:30}). 

Inserting $S_2$ into Eq.~(\ref{eq:13}) and requiring $\Gamma_{\mathsf{b}2J} = 0$, such that $J_{\mathsf{b}}$ and $\psi_{\mathsf{b}}$ are exact conjugate coordinates, gives us
\begin{equation}
g^{\psi}_1  =  \frac{1}{2}\frac{c}{q}\,\eta_{ab}F_b\pd{F_a}{J_{\mathsf{b}0}}-\pd{S_2}{J_{\mathsf{b}0}}.
\end{equation} 

Thus we obtain the bounce-center Lagrangian and Hamiltonian
\begin{eqnarray}
  \label{BCLagrangian}
  \Gamma_{\mathsf{b}}&=&\frac{1}{\epsilon}\frac{q}{c}Y_\alpha dY_\beta+J_\mathsf{b}d\psi_b-K_\mathsf{b}dt,\\
  \mathcal{H}_\mathsf{b} &=& \mathcal{H}_{\mathsf{b}0}  + \otwo,
\end{eqnarray}
where the bounce-center coordinates $(\Y, J_\mathsf{b},\psi_\mathsf{b}; t, K_\mathsf{b})$ are given by 
\begin{eqnarray}
  \label{eq:26}
  Y_a & = & y_a - \epsilon\frac{c}{q}\eta_{ab}F_b+\otwo, \\
  \label{eq:15}
  J_\mathsf{b} & = & J_{\mathsf{b}0}+\epsilon g_1^J + \otwo, \\
  \psi_\mathsf{b} & = & \psi_{\mathsf{b}0} + \epsilon g^{\psi}_1 + \otwo,  \\
  \label{eq:41}
  K_\mathsf{b} &= & K_{\mathsf{g}} - \epsilon F_{t_1} + \otwo, 
\end{eqnarray}
with time $t$ an invariant under the transformation. Note from Eq.~(\ref{eq:26}) that $F_a$ in bounce-center dynamics plays a role similar to the gyroradius vector $\vb{\rho}$ in guiding-center dynamics [Eq.~(\ref{GCCT})]; i.e., $F_a$ may be interpreted as a ``bounce radius'' 2-vector. Also from Eq.~(\ref{eq:41}), $F_{t_1}$ is the oscillatory part of the guiding-center energy coordinate $K_\mathsf{g}$.

The bounce-center equations are then
\begin{eqnarray}
  \dot{Y_a} &=& \epsilon \frac{c}{q}\eta_{ab}\pd{\mathcal{H}_\mathsf{b}}{Y_b}, \\ 
  \dot{\psi_\mathsf{b}} &=& \pd{\mathcal{H}_{\mathsf{b}}}{J_\mathsf{b}}, \\ 
  \dot{J_\mathsf{b}} &=& 0, \\
  \dot{K_{\mathsf{b}}} &=& \pd{\mathcal{H}_{\mathsf{b}}}{t}.
\end{eqnarray}
Thus we see that $J_\mathsf{b}$ is the exact invariant for the bounce motion. 

In Eq.~(\ref{eq:30}),  $\widetilde{g}^J_1=g^{J}_1$ denotes the first-order correction to the second adiabatic invariant $J_{\mathsf{b}0}$. This first-order correction can also be directly obtained from Ref.~\cite{Northrop1963}, where Northrop shows that
\begin{equation}
\td{J_{\bsf 0}}{t} = \omega_{\bsf 0}^{-1}\left[\frac{q}{c}\left( \langle \dot{\alpha}\rangle  \dot{\beta} -  \dot{\alpha} \langle\dot{\beta}\rangle\right) + \left(\dot{K}_\gsf-\langle \dot{K}_\gsf \rangle \right)\right],  
\end{equation}
written using our notation. Since 
\begin{equation}
  \td{J_\bsf}{t} = \td{J_{\bsf 0}}{t} + \epsilon \td{J_{\bsf 1}}{t} +\cdots = 0, 
\end{equation}
and to lowest order, we have $ \langle dJ_{\bsf 0} /dt \rangle = 0$, thus 
\begin{eqnarray}
  \label{eq:52}
  \widetilde{J_{\bsf 1}}&=&-\omega_{\bsf 0}^{-1}\int \td{J_{\bsf 0}}{t}\,d\psi_{\bsf 0}  \nonumber\\ 
                        &=& -\omega_{\bsf 0}^{-2}\int d\psi_{\bsf 0}\left[\frac{q}{c}\left( \langle \dot{\alpha}\rangle  \dot{\beta} -  \dot{\alpha} \langle\dot{\beta}\rangle\right) \right. \nonumber \\
 &  &\left.+\; \left(\dot{K}_\gsf-\langle \dot{K}_\gsf \rangle \right)\right]. 
\end{eqnarray}
Since we have shown that $J_{\bsf 1}$ (i.e., $g^J_1$) is purely oscillatory, we have $J_{\bsf 1} = \widetilde{J_{\bsf 1}}$.  Littlejohn \cite{Littlejohn1982} has shown, for the nonrelativistic case, that the right hand side of Eq.~(\ref{eq:52}) is equal to the right hand side of Eq.~(\ref{eq:30}),  but this result also holds for the relativistic case because the equations have the same functional form. We also note, as pointed out by Littlejohn \cite{Littlejohn1982}, that the Lie-transform approach is more straightforward than the method used in Ref.~\cite{Northrop1966}, which derives the first-order correction to the second adiabatic invariant for nonrelativistic particles in a static magnetic field. 

\section{Hamiltonian theory of drift-center dynamics}
\label{sec:dc}
Starting from the bounce-center Lagrangian (\ref{BCLagrangian}), we now derive the drift-averaged bounce-center Lagrangian, or the drift-center Lagrangian. Similar to the analysis given in Sec.~\ref{sec:bc}, this procedure leads to the first-order correction to the third adiabatic invariant automatically. 

To apply the adiabatic theory to the drift motion, electromagnetic fields must vary on a time scale much slower than the drift period; i.e., $\partial /\partial t \sim \epsilon^2$. We start from the bounce-center Lagrangian (\ref{BCLagrangian}) with term $J_\mathsf{b} d\psi_\mathsf{b}$ dropped, which means we are now  considering a two-dimensional motion parametrized by the constants $J_\mathsf{g}$ and $J_\mathsf{b}$. We set $t_2\equiv\epsilon^2 t$ and the resulting bounce-center Lagrangian is
\begin{equation}
  \label{BCLagrangianIII}
  \Gamma_\mathsf{b}\,=\,\frac{1}{\epsilon}\frac{q}{c}\alpha d\beta-\frac{1}{\epsilon^2}K_\mathsf{b}dt_2\equiv  \frac{1}{\epsilon^2}\left(\epsilon\,\bar{\alpha} d\beta -K_\mathsf{b}dt_2 \right),
\end{equation}
where we henceforth use $\Y=(\alpha, \beta)$ and replaced $q\alpha/c$ by $\bar{\alpha}$ in the last expression.

We now make the usual assumption for the lowest-order motion that in a static field, or with parameter $t_2$ frozen, the orbit of the particle is closed and hence the drift motion of the particle is periodic \cite{Northrop1963}. Thus the coordinates $(\bar{\alpha}, \beta)$ play a role in drift-center dynamics similar to that of as the coordinates $(\ppar, s)$ in bounce-center dynamics. The Hamilton-Jacobi theory again gives us the action-angle variables from canonical coordinates $(\bar{\alpha},\beta)$ as
\begin{equation}
  J_{\mathsf{d}0}(K_\mathsf{b}, t)=\frac{1}{2 \pi}\oint\bar{\alpha} d\beta,
\end{equation}
and $\omega_{\mathsf{d}0}^{-1}=\partial J_{\mathsf{d}0}/\partial {K_\mathsf{b}}$ is the lowest-order angular frequency of the drift motion. Here we use '$\mathsf{d}$' to represent drift motion variables.  The canonically-conjugate coordinate of $J_{\mathsf{d}0}$ is the drift phase  $\psi_{\mathsf{d}0}$. The change from coordinates $(\bar{\alpha},\beta)$ to $(J_{\mathsf{d}0},\psi_{\mathsf{d}0})$ is canonical, thus we have 
\begin{equation}
  \pd{\bar{\alpha}}{J_{\mathsf{d}0}}\pd{\beta}{\psi_{\mathsf{d}0}}-\pd{\bar{\alpha}}{\psi_{\mathsf{d}0}}\pd{\beta}{J_{\mathsf{d}0}}=1,
\end{equation}
which is also valid for the true motion. 

For the lowest-order motion, $J_{\mathsf{d}0}$ is a constant, but with time $t$ unfrozen and higher-order terms included in the true motion, $J_{\mathsf{d}0}$ is no longer an invariant for the drift motion and it will be shown that $d J_{\mathsf{d}0}/dt=\otwo$. Thus we first do a coordinate transformation from $(\bar{\alpha},\beta)$ to $(J_{\mathsf{d}0},\psi_{\mathsf{d}0})$ and then use a Lie transform to construct the true invariant $J_\mathsf{d}$ for the drift motion.  

\subsection{Preliminary coordinate transformation}
Similar to the construction of the bounce-center dynamics in Sec.~\ref{sec:bc}, we first change coordinates from $(\bar{\alpha}, \beta)$ to $(J_{\mathsf{d}0},\psi_{\mathsf{d}0})$. Substituting  
\begin{equation}
    d\beta=\pd{\beta}{J_{\mathsf{d}0}}dJ_{\mathsf{d}0}+\pd{\beta}{\psi_{\mathsf{d}0}}d\psi_{\mathsf{d}0}+\pd{\beta}{t_2}dt_2,
\end{equation}
which is similar to Eq.~(\ref{eq:28}), into Eq.~(\ref{BCLagrangianIII}) gives 
\begin{equation}
  \epsilon^2\Gamma_\mathsf{b}=\epsilon\bar{\alpha}\pd{\beta}{J_{\mathsf{d}0}}dJ_{\mathsf{d}0}+\epsilon\bar{\alpha}\pd{\beta}{\psi_{\mathsf{d}0}}d\psi_{\mathsf{d}0}-\left[K_{\mathsf{b}}-\epsilon \bar{\alpha}\pd{\beta}{t_2}\right]dt_2,
\end{equation}
which gives the lowest- and first-order bounce-center Lagrangians
\begin{eqnarray}
  \Gamma_{\mathsf{b}0}&=&-K_{\mathsf{b}}dt_2, \\
  \Gamma_{\mathsf{b}1}&=&\bar{\alpha}\pd{\beta}{J_{\mathsf{d}0}}dJ_{\mathsf{d}0}+\bar{\alpha}\pd{\beta}{\psi_{\mathsf{d}0}}d\psi_{\mathsf{d}0}+ \bar{\alpha}\pd{\beta}{t_2} dt_2.
\end{eqnarray}

Similar to Eq.~(\ref{eq:16}), we perform a gauge transformation on $\Gamma_{\mathsf{b}1}$; i.e.,  $\Gamma_{\mathsf{b}1}'=\Gamma_{\mathsf{b}1}+dG_{\mathsf{d}}$, such that  
\begin{equation}
  \label{eq:47}
  \Gamma_{\mathsf{b}1J}'=0  \mbox{ and } \Gamma_{\mathsf{b}1\psi}'=J_{\mathsf{d}0},
\end{equation}
where we have again omitted the subscripts of $J_{\mathsf{d}0}$ and $\psi_{\mathsf{d}0}$ when they themselves are subscripts. From Eq.~(\ref{eq:47}), we choose $G_{\mathsf{d}}$ as
\begin{equation}
  \label{eq:17}
  G_{\mathsf{d}}=-\int_0^{J_{\mathsf{d}0}}\Gamma_{\mathsf{b}1J}\, dJ'+f(\psi_{\mathsf{d}0},t_2),
\end{equation}
which is similar to Eq.~(\ref{eq:16}), and $f(\psi_{\mathsf{d}0},t_2)$ is a function that is determined from the condition 
\begin{equation}
  \pd{G_{\mathsf{d}}}{\psi_{{\mathsf{d}0}}}+\Gamma_{\mathsf{b}1\psi} = J_{\mathsf{d}0}.
\end{equation}
Since $\Gamma_{\mathsf{b}1\psi}$ can also be written as
\begin{equation}
  \Gamma_{\mathsf{b}1\psi} = \int^{J_{\mathsf{b}0}}_0 \pd{\Gamma_{\mathsf{b}1\psi}}{J'} dJ' + \left.\Gamma_{\mathsf{b}1\psi}\right\vert_{J_{\mathsf{b}0}=0}, 
\end{equation}
the equation for $f(\psi_{\mathsf{d}0},t_2)$ becomes
\begin{equation}
  \pd{f}{\psi_{\mathsf{d}0}}=-\left.\bar{\alpha}\pd{\beta}{\psi_{\mathsf{d}0}}\right|_{J_{\mathsf{b}0}=0}. 
\end{equation}
Note the difference between Eqs.~(\ref{eq:17}) and (\ref{eq:16}), since we generally do not have $ \left.\Gamma_{\mathsf{b}1\psi}\right\vert_{J_{\mathsf{b}0}=0}= 0$. 
Finally the gauge transformation (\ref{eq:17}) yields the new $t_2$-term 
\begin{eqnarray}
  \Gamma_{\mathsf{b}1t_2}' & = & -\int^{J_{\mathsf{d}0}}_0\pd{\Gamma_{\mathsf{b}1J}}{t_2}dJ'+\pd{f}{t_2} + \bar{\alpha}\pd{\beta}{t_2}\nonumber \\
 & \equiv & F_{t_2}(J_{\mathsf{d}0},\psi_{\mathsf{d}0},t_2) .
\end{eqnarray}
Now we have our zeroth- and first-order bounce-center Lagrangian 
\begin{eqnarray}
  \label{eq:18}
  \Gamma_{\mathsf{b}0} &=& -K_{\mathsf{b}}dt_2, \\
  \label{eq:19}
  \Gamma_{\mathsf{b}1} &=& J_{\mathsf{d}0} d\psi_{\mathsf{d}0}+F_{t_2} dt_2,
\end{eqnarray}
where we have dropped the prime, with the extended Hamiltonian 
\begin{equation}
  \label{eq:20}
  \mathcal{H}_{\mathsf{b}0} = H_{\mathsf{b}0} - K_\mathsf{b}.
\end{equation}

From $\Gamma_{\mathsf{b}0}$ and $\Gamma_{\mathsf{b}1}$, we obtain the Euler-Lagrange equation for $J_{\mathsf{d}0}$ 
\begin{equation}
  \label{driftphi}
  \td{J_{\mathsf{d}0}}{t}=\epsilon^2\pd{F_{t_2}}{\psi_{\mathsf{d}0}}+\mathcal{O}(\epsilon^3).
\end{equation}

Comparing Eqs.~(\ref{eq:18})-(\ref{eq:20}) with Eqs.~(\ref{eq:21})-(\ref{eq:23}), we find that the bounce-center and guiding-center equations are very similar, except that in Eq.~(\ref{eq:18}) we do not have the $d\beta$ term and in Eq.~(\ref{eq:19}) we do not have the $dy_a$ term. Thus the Lie transform from the bounce-center coordinates to the drift-center coordinates will be very similar to the Lie transform from the guiding-center coordinates to the bounce-center coordinates.  

\subsection{Lie Transform from $(J_{\mathsf{d}0},\psi_{\mathsf{d}0}; t, K_\mathsf{b})$ to $(J_\mathsf{d},\psi_\mathsf{d}; t, K_\mathsf{d})$}
The lowest-order drift-center Lagrangian and Hamiltonian are given by $\Gamma_{\mathsf{d}0} = \Gamma_{\mathsf{b}0}$ and $\mathcal{H}_{\mathsf{d}0} = \mathcal{H}_{\mathsf{b}0}$. The first-order Lagrangian is given by 
\begin{equation}
  \label{eq:48}
\Gamma_{\mathsf{d}1} = \Gamma_{\mathsf{b}1} - i_1\cdot\Omega_{\mathsf{b}0} + dS_1.
\end{equation}
Choosing $S_1 = 0$ and substituting $i_1\cdot\Omega_{\mathsf{b}0} = -g_1^K dt_2$ and  $\Gamma_{\mathsf{b}1}$ from Eq.~(\ref{eq:19}) into Eq.~(\ref{eq:48}), we obtain 
\begin{equation}
  \label{eq:49}
  \Gamma_{\mathsf{d}1} = J_{\mathsf{d}0} d\psi_{\mathsf{d}0} +  (F_{t_2} + g_1^K)dt_2.
\end{equation}
The first-order Hamiltonian is then given by 
\begin{equation}
  \label{eq:50}
  \mathcal{H}_{\mathsf{d}1} = \mathcal{H}_{\mathsf{b}1} - g_1\cdot d\mathcal{H}_{\mathsf{b}0} = -g_1^J \omega_{\mathsf{d}0} + g_1^K,
\end{equation}
where we have used $\mathcal{H}_{\mathsf{b}1} = 0$.  Note that Eqs.~(\ref{eq:49}) - (\ref{eq:50}) look similar to Eqs.~(\ref{eq:31}) and (\ref{eq:32}) and we omit the remaining details here.

The first-order coordinate generators from the above Lie transform are 
\begin{eqnarray}
  g^J_1 &\equiv& \pd{S_2}{\psi_{\mathsf{d}0}} = -\omega_{\mathsf{d}0}^{-1}\widetilde{F_{t_2}}, \\
  g^\psi_1 & = & -\pd{S_2}{J_{\mathsf{d}0}}, \\
  g^K_1 & = & -F_{t_2}, 
\end{eqnarray}
where a tilde in this Section indicates the drift-phase oscillatory part. Thus the overall coordinate transformation is   
\begin{eqnarray}
  \label{driftJ}
  J_\mathsf{d}&=&J_{\mathsf{d}0}-\epsilon\, \omega_{\mathsf{d}0}^{-1}\widetilde{F_{t_2}}+\mathcal{O}(\epsilon^2), \\
  \psi_\mathsf{d}&=&\psi_{\mathsf{d}0}+\epsilon g^{\psi}_1+\mathcal{O}(\epsilon^2), \\
  K_\dsf &=& K_\bsf - \epsilon F_{t_2} + \otwo. 
\end{eqnarray}

The drift-center Lagrangian written in coordinates $(J_\mathsf{d},\psi_\mathsf{d}; t, K_{\mathsf{d}})$ is
\begin{equation}
  \Gamma_\mathsf{d}=\frac{1}{\epsilon}\,J_\mathsf{d}\,d\psi_\mathsf{d}-K_\mathsf{d} dt,
\end{equation}
and the drift-center Hamiltonian function is 
\begin{equation}
  \mathcal{H}_\mathsf{d} = \mathcal{H}_{\mathsf{d}0}+\epsilon \mathcal{H}_{\mathsf{d}1}+\otwo,
\end{equation}
where $\mathcal{H}_{\mathsf{d}0} = \mathcal{H}_{\mathsf{b}0}$ and $\mathcal{H}_{\mathsf{d}1} = -\langle\langle F_{t_2} \rangle\rangle$, with a drift-phase average denoted as $\langle\langle \cdots \rangle\rangle$. 

The drift-center equations of motion are obtained from the Euler-Lagrange equations
\begin{eqnarray}
  \label{eq:33}
  \dot{J_\mathsf{d}} &=& 0,  \\
  \dot{\psi_\mathsf{d}} &=& \pd{\mathcal{H}_\mathsf{d}}{J_\mathsf{d}}, \\
  \dot{K_\mathsf{d}} &=& \pd{\mathcal{H}_\mathsf{d}}{t}.
\end{eqnarray}

Equation (\ref{driftJ}) shows the first-order correction to the third adiabatic invariant. We can also write the oscillatory part of $F_{t_2}$ in another form by using Eq.~(\ref{driftphi}) and 
\begin{equation}
  \label{eq:1}
  \td{J_{\mathsf{d}0}}{t}=\pd{J_{\mathsf{d}0}}{t}+\pd{J_{\mathsf{d}0}}{K_{\mathsf{b}}}\dot{K}_{\mathsf{b}},
\end{equation}
which gives that 
\begin{equation}
  \widetilde{F_{t_2}}=\int d\psi_{\mathsf{d}0}'\left(\pd{J_{\mathsf{d}0}}{t}+\frac{1}{\omega_{\mathsf{d}0}}\dot{K}_{\mathsf{b}}\right),
\end{equation}
where we have set $\epsilon = 1$.  Northrop \cite{Northrop1963} has shown that the first term on the right side of Eq.~(\ref{eq:1}) can also be written as 
\begin{equation}
  \pd{J_{\mathsf{d}0}}{t} = -\omega_{\dsf 0}^{-1}\langle\langle \dot{K_\bsf} \rangle\rangle. 
\end{equation}
With $\partial J_{\dsf 0} / \partial K_\bsf = \omega_{\dsf 0}^{-1}$, we find 
\begin{equation}
  \widetilde{F_{t_2}}=\omega_{\dsf 0}^{-1}\int d\psi_{\mathsf{d}0}'\left(\dot{K}_\bsf - \langle\langle \dot{K_\bsf} \rangle\rangle\right).
\end{equation} 
Thus, Ref.~\cite{Northrop1963} implicitly contains the first-order correction term to the third adiabatic invariant (see Eq.~(3.80) on page 64 of Ref.~\cite{Northrop1963}), but Eq.~(\ref{driftJ}) is an explicit expression.  

\section{Summary and Discussion}
\label{sec:summary}
In this work, we have presented the Hamiltonian theory of adiabatic motion of a relativistic charged particle and the derivation of the first-order corrections to the second and third adiabatic invariants. The background electromagnetic fields vary on the drift time scale when we consider the guiding-center motion and the bounce-center motion.  The effect of these time-varying background fields on the guiding-center motion is shown by the extra terms in the guiding-center equations (\ref{dX/dt}) and (\ref{dppar/dt}), compared to the guiding-center equations in Ref.~\cite{Brizard1999}.  The first-order correction to the second adiabatic invariant of a relativistic particle is then shown in Eq.~(\ref{eq:15}). To apply the adiabatic analysis to the drift motion, we assume that the background fields vary on a time scale much smaller than the drift period.  The first-order correction to the third adiabatic invariant is shown in Eq.~(\ref{driftJ}). 

This work simplifies previous work on relativistic guiding-center motion, generalizes previous work on bounce-center motion for a relativistic particle in time-varying fields, and extends previous work on drift-center motion using Lie-transform perturbation methods in extended phase space. These results are especially useful in space plasma physics, where adiabatic theory is the foundation for modeling and understanding the dynamics of magnetically-trapped energetic particles.

The hierarchy of the adiabatic motions in this work may be shown as follows
\begin{widetext}
\begin{equation}
  \nonumber
  (\x, \p; t, W_\mathsf{p}) \xrightarrow{\mbox{\hspace{0.15cm} \gsf\hspace{0.15cm}}}
  \left\{
  \begin{array}{l}
    (\X, \ppar; t, W_\mathsf{g})   \xrightarrow{\mbox{\hspace{0.15cm} \bsf\hspace{0.15cm}}}
    \left\{
     \begin{array}{l}
       (\alpha, \beta;  t, K_\mathsf{b})  \xrightarrow{\mbox{\hspace{0.15cm} \dsf\hspace{0.15cm}}}
       \left\{
       \begin{array}{l}
	 (t, K_\mathsf{d}) \\
	 (J_\mathsf{d}, \psi_\mathsf{d}) 
       \end{array}
       \right.
       \\
       (J_{\mathsf{b}}, \psi_{\mathsf{b}})
     \end{array}    \right.
     \\
    (J_\mathsf{g}, \psi_{\mathsf{g}})
  \end{array}
  \right.
\end{equation}
\end{widetext}
where $J_{\mathsf{g}}$ is related to the first invariant $\mu$ by $J_{\mathsf{g}} = (mq/c)\,\mu$ and $\psi_{\mathsf{g}}=\theta$. The first arrow (\gsf) thus indicates the gyro-phase average process, the second arrow (\bsf) the bounce-phase average and the third arrow (\dsf) the drift-phase average. 

In this paper we have shown how first-order corrections to adiabatic invariants can be obtained using Lie-transform methods. Alternatively, the oscillatory part of the first-order correction to an adiabatic invariant can be obtained as follows. Differenting the exact invariant
\begin{equation}
  J_\ksf = J_{\ksf 0} + \epsilon J_{\ksf 1} + \cdots,
\end{equation}
where $\ksf$ can be $\gsf, \bsf$ or $\dsf$, to lowest order gives
\begin{equation}
  \td{J_\ksf}{t} = \td{J_{\ksf 0}}{t} + \epsilon\, \omega_{\ksf 0} \td{J_{\ksf 1}}{\psi_{\ksf 0}} + \cdots = 0.
\end{equation}
Since $J_{\ksf 0}$ satisfies the necessary condition 
\begin{equation}
  \left\langle \td{J_{\ksf 0}}{t} \right\rangle_\ksf = 0,
\end{equation}
where $\langle\cdots\rangle_\ksf$ denotes the fast-angle average canonically conjugate to $J_{\ksf}$, and we obtain the oscillatory part of $J_{\ksf 1}$ as
\begin{equation}
  \epsilon \widetilde{J_{\ksf 1}} = -\omega_{\ksf 0}^{-1}\int \td{J_{\ksf 0}}{t} d\psi_{\ksf 0}.
\end{equation}
The phase-independent part of $J_{\ksf 1}$ can be obtained by using the Lie-transform method. 

The use of Hamiltonian theory in describing adiabatic motions results in equations that satisfy energy conservation for time-independent fields and preserve phase-space volume naturally, in contrast to the results of Northrop \cite{Northrop1963}. These conservation laws are very useful for checking numerical accuracy in simulations. Based on this work, fluctuations of electromagnetic fields can be added to the background fields and equations of motion in the presence of electromagnetic waves can be derived, as in Refs.~\cite{Brizard1999}, \cite{Brizard2000}, and \cite{Brizard2004}. 

\acknowledgments

The Authors wish to acknowledge the anonymous referee for pointing out the work of Grebogi and Littlejohn \cite{Grebogi1984}. This work was supported by NSF grants ATM-0316195, and ATM-0000950,  and NASA grants NNG05GH93G and NNG05GJ95G.

\end{document}